\documentclass[
  reprint,
  aps,
  prl,
  floatfix,
  raggedbottom,
  superscriptaddress,
  nobibnotes,
  amsmath,amssymb,
]{revtex4-2}

\usepackage{graphicx,stackengine}
\usepackage{xcolor}
\usepackage{bm}
\usepackage{mathrsfs}
\usepackage{tikz}
\usetikzlibrary{arrows.meta,positioning}
\graphicspath{{../figures/}{figures/}}

\newcommand{\dd}{\mathrm{d}}
\newcommand{\ee}{\mathrm{e}}
\newcommand{\RR}{\mathbb{R}}

\newcommand{\PP}{\mathcal{P}}

\newcommand{\LLop}{\mathcal{L}}
\newcommand{\Jloc}{J^{\mathrm{loc}}}
\newcommand{\Jnl}{J^{\mathrm{nl}}}
\newcommand{\stot}{s_{\mathrm{tot}}}
\newcommand{\Vss}{V^{\mathrm{ss}}}
\newcommand{\srel}{s_{\mathrm{rel}}}
\newcommand{\sact}{s_{\mathrm{act}}}

\begin{document}

\title{Active Jarzynski and Crooks Relations for Driven Jump-Diffusions}

\author{Shuyuan Fan}
\email{fanshy9@mail2.sysu.edu.cn}
\email{233621002@stu.gbu.edu.cn}
\affiliation{School of Mathematics, Sun Yat-sen University, Guangzhou, Guangdong 510275, China}
\affiliation{School of Science, Great Bay University, Dongguan, Guangdong 523000, China}

\author{Bing Miao}
\email{bmiao@ucas.ac.cn}
\affiliation{Center of Materials Science and Optoelectronics Engineering, College of Materials Science and Opto-Electronic Technology, University of Chinese Academy of Sciences (UCAS), Beijing 100049, China}

\author{Jinqiao Duan}
\email{duan@gbu.edu.cn}
\affiliation{School of Science, Great Bay University, Dongguan, Guangdong 523000, China}

\author{Qi Zhang}
\thanks{\href{mailto:qzhang@bimsa.cn}{qzhang@bimsa.cn} (corresponding author)}
\affiliation{Beijing Institute of Mathematical Sciences and Applications, Beijing 101408, China}

\date{\today}

\begin{abstract}

  Non-conservative active driving breaks detailed balance in non-Gaussian jump-diffusions, obscuring the connection between external protocol work, housekeeping dissipation, and free-energy differences. Here, we establish exact finite-time fluctuation relations for driven Markov jump-diffusions—encompassing infinite-activity Lévy processes—by comparing path probabilities under the forward and current-reversed dual dynamics relative to the instantaneous steady state (ISS). We derive a trajectory-level decomposition of total entropy production that cleanly separates transient state relaxation, housekeeping heat, and external protocol driving. When the ISS takes a Gibbsian form, gauge-tilting the dual path measure yields an Active Jarzynski equality, $\bigl\langle\exp[-(W+\Delta s_{\rm act})]\bigr\rangle_{\pi_0}=e^{-\Delta F}$, which incorporates active housekeeping entropy $\Delta s_{\rm act}$ alongside mechanical work $W$ and coexists with the conventional Jarzynski equality. Under protocol reversal, this dual framework further reveals an Active Crooks relation, where the asymmetry between forward and reverse work distributions is governed by the conditional exponential average of active dissipation. Our results provide a framework for recovering free-energy differences in driven jump-diffusions with Gibbsian instantaneous steady states and for characterizing active irreversibility through trajectory statistics.
\end{abstract}

\maketitle

\paragraph*{Introduction.}

Quantifying how external driving and internal fluctuations govern thermodynamic irreversibility is a central theme of modern statistical physics~\cite{Seifert2012,Gallavotti1995}. In classical stochastic thermodynamics, a system coupled to a thermal bath (in thermal units $k_{\rm B}T_{\rm bath} = 1$) and driven from an initial equilibrium state $\pi_0$ is assumed to obey microscopic reversibility~\cite{Sekimoto1998,Seifert2005}. When detailed balance holds at each frozen value of an external protocol, this time-reversal symmetry underpins two foundational results: the Jarzynski equality~\cite{Jarzynski1997PRL}, $\langle \ee^{-W}\rangle_{\pi_0} = \ee^{-\Delta F}$, which determines the equilibrium free-energy difference $\Delta F$ from the exponential average of the protocol work $W$, and the Crooks fluctuation theorem~\cite{Crooks1999}, $p_F(w)/p_R(-w) = \ee^{w-\Delta F}$, which compares the probability density of forward work $w$ with that of reverse work $-w$.

This classical paradigm does not extend directly to active systems undergoing discontinuous, non-Gaussian transitions~\cite{Szamel2019,PietzonkaSeifert2018}. Across diverse scales—ranging from self-propelled colloids and molecular motors to transcriptional bursting circuits~\cite{Golding2005}—energy-consuming processes can generate non-conservative drift forces and jump noise. Driving such active jump-diffusions with a time-dependent protocol introduces three interlocking challenges.
First, the system generally lacks a common time-independent steady state to serve as a baseline under a time-dependent protocol~\cite{ZhangLu2025,FanZhang2026}. Second, mechanical protocol work $W$ must be distinguished from continuous and jump-induced active dissipation (housekeeping entropy $\Delta s_{\rm act}$). Third, trajectory comparisons must account for jump-rate asymmetry and the associated compensators; forbidden reverse jumps can make forward and reversed path measures singular. Together, these effects disrupt standard work fluctuation symmetries, confounding external mechanical manipulation with autonomous active driving~\cite{Szamel2023,Dabelow2019}.

\begin{figure}[htbp]
  \centering
  \resizebox{0.98\columnwidth}{!}{%
    \begin{tikzpicture}[
      >=Latex, line width=0.45pt,
      box/.style={draw, rounded corners=2pt, align=center, inner sep=5pt,
        font=\small},
      lab/.style={font=\small, align=center}
    ]
      \node[box] (P) at (0,0) {$\PP$\\forward};
      \node[box] (Pa) at (4,0) {$\PP^\alpha$\\reference};
      \node[box] (Pg) at (8,0) {$\PP^{\alpha,\Gamma}$\\weighted reference};
      \draw[->] (P) -- node[lab, above] {Conjugate}
        node[lab, below] {dynamics} (Pa);
      \draw[->] (Pa) -- node[lab, above] {Gauge tilt}
        node[lab, below] {$e^{-\Gamma}/Z_\alpha^\Gamma$} (Pg);
      \node[box, anchor=north, text width=4.2cm, minimum height=3.4cm]
        (phys) at (1.4,-2.5)
        {Path functionals\\[5pt]
          $\displaystyle\mathcal I^\alpha=
            \log\frac{\dd\PP}{\dd\PP^\alpha}$\\[6pt]
          $\mathcal I^\Theta,\ \Delta\stot$\\[3pt]
          $\Delta\sact$};
      \node[box, anchor=north, text width=4.2cm, minimum height=3.4cm]
        (ft) at (7,-2.5)
        {Fluctuation relations\\[5pt]
          $\displaystyle\Delta s^{\alpha,\Gamma}=
            \log\frac{\dd\PP}{\dd\PP^{\alpha,\Gamma}}$\\[5pt]
          $\langle e^{-\Delta s^{\alpha,\Gamma}}\rangle_{\PP}=1$\\[5pt]
          \mbox{(Active) Jarzynski},\\
          \mbox{(Active) Crooks}};
      \draw[->] (P.south) -- ([xshift=-8mm]phys.north);
      \draw[->] (Pa.south) -- ([xshift=8mm]phys.north);
      \draw[->] (Pg.south) -- ([xshift=8mm]ft.north);
      \draw[->, preaction={draw=white, line width=2.5pt,
        shorten >=4pt, shorten <=4pt}]
        (P.south) -- ([xshift=-8mm]ft.north);
    \end{tikzpicture}
  }
  \vspace{3pt}
  \caption{Reference comparisons and fluctuation relations. The left pair of arrows compares $\PP$ with $\PP^\alpha$ to identify path functionals. The right pair compares $\PP$ with $\PP^{\alpha,\Gamma}$ to obtain the weighted IFT. The upper arrows select conjugate dynamics and then apply the gauge weight.}
  \label{fig:path_measure_geometry}
\end{figure}
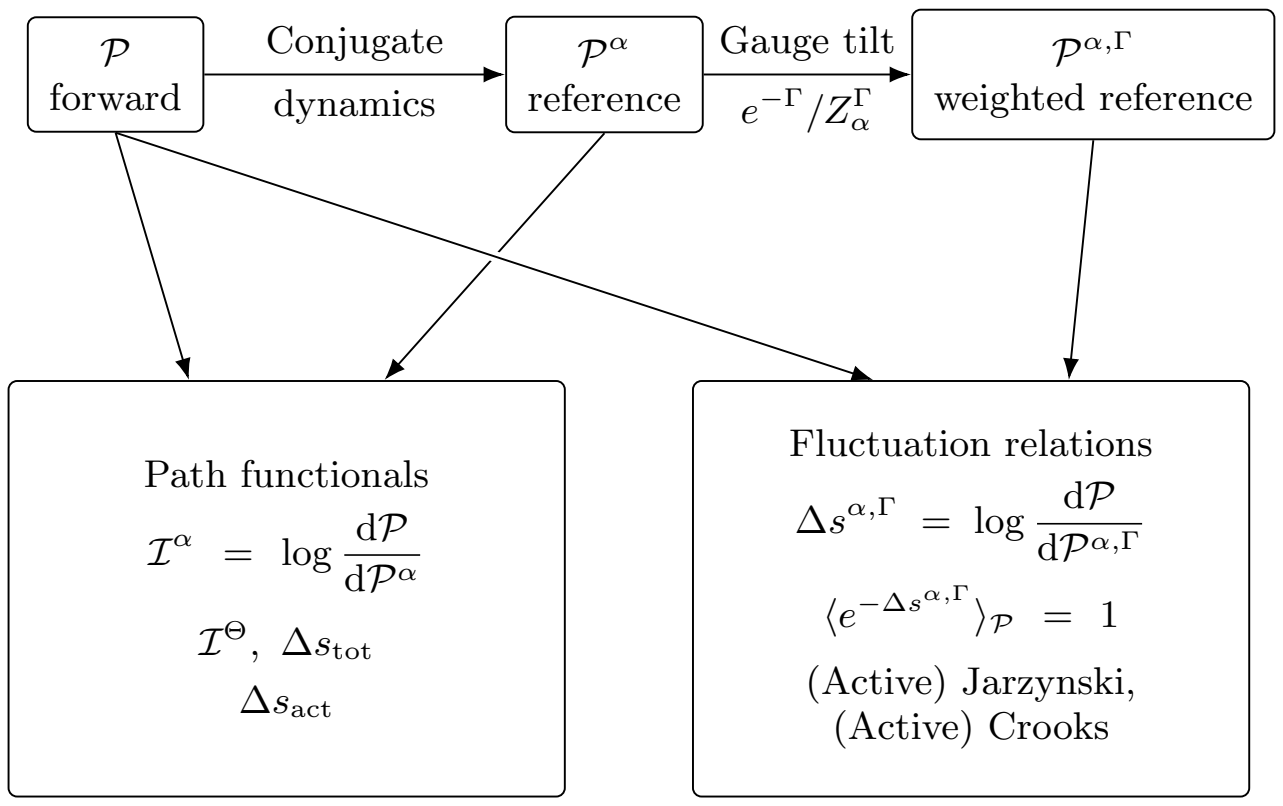

To address these difficulties, we introduce the \emph{instantaneous steady state} (ISS) $\pi_t$ at each frozen protocol value and use it as a moving thermodynamic reference~\cite{HatanoSasa2001}. This preserves the familiar distinction between relaxation toward a stationary density and the dissipation sustained by its currents, while accounting for the changing protocol. To connect this thermodynamic separation to trajectory probabilities, we compare the original process with dual dynamics that preserve the same frozen density but reverse its stationary currents~\cite{ZhangLu2025}.
For a common initial distribution, the logarithmic ratio of the original and dual path probabilities gives the active entropy $\Delta s_{\rm act}$. Weighting the original trajectories by $e^{-\Delta s_{\rm act}}$ therefore gives the dual path measure.  
Combining these dual dynamics with an exponential gauge transformation with the work weight $e^{-W}$ (Fig.~\ref{fig:path_measure_geometry}) then enables us to reweight trajectory ensembles by mechanical work and establish exact finite-time work relations.

This dual-gauge framework unifies non-equilibrium thermodynamics at the ensemble and trajectory levels. When the frozen ISS admits a Gibbsian form $\pi_t(x) = Z_t^{-1}\ee^{-H(t,x)}$ for a prescribed, protocol-controlled Hamiltonian $H(t,x)\equiv H(\lambda_t,x)$, the macroscopic entropy production rate $\dot S_{\mathrm{tot}}(t)$ obeys the decomposition
\begin{equation}\label{eq:macro_balance_intro}
    \dot S_{\mathrm{tot}}(t) = \dot{\overline W}(t) + \dot{Q}_{\rm hk}(t) - \dot{\mathcal{F}}_t - \frac{\dd}{\dd t}D(\rho_t\Vert\pi_t) \ge 0,
\end{equation}
where $\dot{\overline W}$ and $\dot Q_{\rm hk}$ denote the mean protocol power and housekeeping heat rate~\cite{FanZhang2026}, $\overline W(t):=\langle W[0,t]\rangle_{\PP}$ is the mean work, $\mathcal F_t = -\log Z_t$ is the equilibrium free energy, and the Kullback--Leibler divergence $D(\rho_t\Vert\pi_t)$ quantifies the lag of the evolving density $\rho_t$ behind $\pi_t$. At the trajectory level, total entropy production $\Delta\stot$ accumulated over $[0,T]$ decomposes into
\begin{equation}
    \Delta\stot = W + \Delta \sact - \Delta F + \Delta \srel,
  \label{eq:traj_decomp_intro}
\end{equation}
where $\Delta F=\mathcal F_T-\mathcal F_0=-\log(Z_T/Z_0)$ is the equilibrium free-energy difference, $\Delta \sact$ is the trajectory-level active housekeeping heat, $W$ is the mechanical work, and the endpoint relaxation term $\Delta \srel$ is related to the Kullback--Leibler divergence above through
\begin{equation*}
    \bigl\langle\Delta\srel\bigr\rangle_{\PP}
    =D(\rho_0\Vert\pi_0)-D(\rho_T\Vert\pi_T).
\end{equation*}
Crucially, comparing the forward path measure with the work-tilted dual path measure yields the finite-time \emph{Active Jarzynski equality},
\begin{equation}
    \left\langle \ee^{-(W+\Delta s_{\rm act})}\right\rangle_{\pi_0} = \ee^{-\Delta F}.
  \label{eq:active_jarzynski_intro}
\end{equation}
Comparing forward paths with reversed records of the protocol-reversed experiment, initialized in $\pi_T$, establishes the \emph{Active Crooks relation},
\begin{equation}
    \frac{p_F(w)}{p_R(-w)} = \frac{\ee^{w-\Delta F}}{\left\langle\ee^{-\Delta \sact} \,\middle|\, W=w\right\rangle_{\pi_0}},
  \label{eq:active_crooks_intro}
\end{equation}
where the conditional active factor describes the active contribution to the asymmetry between forward work $w$ and reverse work $-w$.

\paragraph*{Jump-diffusion dynamics and gauge-tilted path measures.}

We consider a driven jump-diffusion $X_t \in \RR^d$ on $[0,T]$ governed by the It\^o equation
\begin{equation}
    \dd X_t = b_t(X_t)\dd t + \sqrt{2}\,a\,\dd W_t + \int_{\RR^d}(y-X_{t^-})N_X(\dd t,\dd y),
  \label{eq:sde_main}
\end{equation}
where $b_t(x)$ is the time-dependent drift, $W_t$ is a standard $d$-dimensional Wiener process, $A = aa^{\mathsf{T}} > 0$ is a constant diffusion tensor characterizing thermal fluctuations from a bath at temperature $T_{\rm bath}$, and $N_X(\dd t,\dd y)$ is the jump measure generated by Poisson noise, with state-dependent compensator $q_t(X_{t^-},y)\dd y\dd t$. The transition rate kernel $q_t(x,y)$ governs non-local jumps from $x$ to $y$ (e.g., transcriptional bursts~\cite{Golding2005}), encompassing infinite-activity L\'evy processes with summable small jumps.

Let $\PP$ denote the path probability measure of trajectories $X = \{X_t\}_{t \in [0,T]}$ initialized in $\rho_0$, with instantaneous density $\rho_t(x)$. An auxiliary reference process with drift $b_t^\alpha$, jump kernel $q_t^\alpha$, and initial density $\rho_0^\alpha$ has path measure $\PP^\alpha$. The trajectory log-likelihood ratio between $\PP$ and $\PP^\alpha$ is
\begin{equation}
    \mathcal{I}^\alpha[X] = \log\frac{\dd\PP}{\dd\PP^\alpha}[X].
  \label{eq:relative_om_def}
\end{equation}
Averaging the jump-diffusion Girsanov formula~\cite{CheriditoFilipovicYor2005} [End Matter, Eq.~\eqref{eq:em-relative-entropy}] gives the path-space relative entropy $\langle \mathcal{I}^\alpha \rangle_\PP = D(\PP\Vert\PP^\alpha) \ge 0$:
\begin{align}
    & \langle\mathcal I^\alpha\rangle_\PP \nonumber \\
    =&D(\rho_0\Vert\rho_0^\alpha)
    +\frac14\int_0^T\!\!\int\rho_t (b_t-b_t^\alpha)^{\mathsf T}A^{-1} (b_t-b_t^\alpha) \dd x\dd t\nonumber\\
    &+\int_0^T\!\!\iint\rho_t(x)
    \left[q_t\log\frac{q_t}{q_t^\alpha}-q_t+q_t^\alpha\right]
    \dd y\dd x\dd t,
  \label{eq:om_mean_main}
\end{align}
where the three terms correspond to initial distribution mismatch, drift mismatch, and jump-rate asymmetry, with $-q_t+q_t^\alpha$ accounting for the escape-rate (traffic) correction~\cite{Maes2021}.

To incorporate thermodynamic observables such as mechanical work, we apply an exponential gauge tilt to $\PP^\alpha$ with an additive trajectory functional $\Gamma[X]$. For a finite path-space partition function $Z_\alpha^\Gamma = \langle \ee^{-\Gamma}\rangle_{\PP^\alpha}$, the tilted measure satisfies
\begin{equation}
    \frac{\dd\PP^{\alpha,\Gamma}}{\dd\PP^\alpha} = \frac{\ee^{-\Gamma}}{Z_\alpha^\Gamma}.
  \label{eq:gauge_path_measure}
\end{equation}
By the chain rule for Radon--Nikodym derivatives,
\begin{equation}
    \frac{\dd\PP^{\alpha,\Gamma}}{\dd\PP} = \frac{\dd\PP^{\alpha,\Gamma}}{\dd\PP^\alpha}\frac{\dd\PP^\alpha}{\dd\PP} = \frac{\ee^{-(\mathcal{I}^\alpha + \Gamma)}}{Z_\alpha^\Gamma}.
  \label{eq:gauge_chain_main}
\end{equation}
Normalization of $\PP^{\alpha,\Gamma}$ immediately yields the master integral fluctuation theorem (IFT),
\begin{equation}
    \left\langle \ee^{-\Delta s^{\alpha,\Gamma}} \right\rangle_\PP = 1,
    \qquad
    \Delta s^{\alpha,\Gamma} = \mathcal{I}^\alpha + \Gamma + \log Z_\alpha^\Gamma,
  \label{eq:master_ift}
\end{equation}
which implies the generalized second-law inequality $\langle\Delta s^{\alpha,\Gamma}\rangle_\PP \ge 0$ via Jensen's inequality.

Because full trajectory functionals like $\mathcal{I}^\alpha$ are often less accessible experimentally than cumulative energy inputs, we also compare the marginal distributions of a trajectory observable $\mathcal{O}[X]$ (e.g., protocol work $W$). Conditioning the likelihood ratio $\dd\PP^\alpha/\dd\PP = \ee^{-\mathcal{I}^\alpha}$ on any measurable event $\mathcal{O}\in B$ (with $\PP\{\mathcal{O}\in B\} > 0$) gives the conditional relation
\begin{equation}
    \frac{\PP^\alpha\{\mathcal{O}\in B\}}{\PP\{\mathcal{O}\in B\}} = \left\langle\ee^{-\mathcal{I}^\alpha} \,\middle|\, \mathcal{O}\in B\right\rangle_\PP.
  \label{eq:conditional_observable}
\end{equation}
Selecting physically motivated reference dynamics $\PP^\alpha$ and gauge weights $\Gamma$ in Eqs.~\eqref{eq:master_ift} and \eqref{eq:conditional_observable} allows us to systematically derive both integral and detailed work relations.

\paragraph*{Macroscopic free-energy dissipation.}
The density $\rho_t(x)$ satisfies the Fokker--Planck equation $\partial_t\rho_t=\LLop_t^*\rho_t$, where $\LLop_t^*$ is the adjoint of the jump-diffusion generator $\LLop_t$. At each frozen time $t$, the instantaneous steady state (ISS) $\pi_t$ satisfies $\LLop_t^*\pi_t=0$. Using $\pi_t$ as a time-dependent reference, the macroscopic total entropy production rate $\dot S_{\rm tot}(t)$ decomposes into nonadiabatic and active housekeeping contributions~\cite{Esposito2010,FanZhang2026} [see Eqs.~\eqref{eq:em-total-rate}--\eqref{eq:em-na-rate} in the End Matter for explicit current--affinity expressions]:
\begin{align}\label{eq:macroscopic_split_main}
    \dot S_{\mathrm{tot}}(t) &= -\frac{\dd}{\dd t} D(\rho_t \Vert \pi_t) + \frac{\dd}{\dd t}\langle Y_{\rm HS}[0,t]\rangle + \dot S_{\mathrm{act}}(t) \nonumber \\
    &= \dot S_{\rm na}(t)+ \dot S_{\rm act}(t) \ge 0.
\end{align}
Here, $D(\rho_t\Vert\pi_t) = \int \rho_t\log(\rho_t/\pi_t)\dd x$ is the relative entropy, $Y_{\rm HS}[0,t]=-\int_0^t(\partial_s\log\pi_s)(X_s)\dd s$ is the Hatano--Sasa protocol functional~\cite{HatanoSasa2001} with driving rate $\frac{\dd}{\dd t}\langle Y_{\rm HS}[0,t]\rangle = -\int\rho_t\partial_t\log\pi_t\,\dd x$, and
\begin{equation}\label{eq:nonadiabatic_rate_main}
    \dot S_{\mathrm{na}}(t) = -\frac{\dd}{\dd t}D(\rho_t\Vert\pi_t) + \frac{\dd}{\dd t}\langle Y_{\rm HS}[0,t]\rangle \ge 0
\end{equation}
is the nonadiabatic entropy production rate, while $\dot S_{\rm act}(t) = \dot{Q}_{\mathrm{hk}}(t) \ge 0$ denotes the active housekeeping dissipation rate sustaining steady-state currents.

When the ISS admits a Gibbsian form $\pi_t(x)=Z_t^{-1}\ee^{-H(t,x)}$ with partition function $Z_t=\int \ee^{-H(t,x)}\dd x$ and equilibrium free energy $\mathcal F_t=-\log Z_t$, the nonequilibrium free energy $\mathcal F_{\rm neq}(t) = \int\rho_t [H(t,x) + \log\rho_t]\dd x$ satisfies $\mathcal F_{\rm neq}(t) - \mathcal F_t = D(\rho_t\Vert\pi_t)$. In this Gibbs--ISS setting, the protocol driving rate becomes $\frac{\dd}{\dd t}\langle Y_{\mathrm{HS}}[0,t]\rangle = \dot{\overline W}(t) - \dot{\mathcal{F}}_t$, where $\dot{\overline W}(t) = \int \rho_t(x) \partial_t H(t,x)\dd x$ is the mean mechanical power. Consequently, Eq.~\eqref{eq:nonadiabatic_rate_main} implies $\dot{\mathcal{F}}_{\mathrm{neq}}(t) = \dot{\overline W}(t) - \dot S_{\mathrm{na}}(t)$, and Eq.~\eqref{eq:macroscopic_split_main} yields $\dot{\mathcal{F}}_{\mathrm{neq}}(t) = \dot{\overline W}(t) + \dot{Q}_{\mathrm{hk}}(t) - \dot S_{\mathrm{tot}}(t)$, directly recovering the dynamic balance in Eq.~\eqref{eq:macro_balance_intro}. For a static protocol ($\dot{\overline W}=\dot{\mathcal{F}}_t=0$), $Y_{\rm HS}$ vanishes and $\dot S_{\rm na}(t) = -\frac{\dd}{\dd t}\mathcal{F}_{\rm neq}(t)$ captures purely transient free-energy relaxation; once the system reaches a nonequilibrium steady state $\pi$ breaking detailed balance, this relaxation ceases ($\dot S_{\rm na}=0$) while housekeeping heat persists ($\dot S_{\rm tot}=\dot S_{\rm act}>0$).

\paragraph*{Reference dynamics and entropy relations.}

Along an individual trajectory, the total entropy production $\Delta\stot[0,T]=\Delta s_{\rm sys}+Q[0,T]$ combines the system-entropy change $\Delta s_{\rm sys}=-\log[\rho_T(X_T)/\rho_0(X_0)]$ with the environment-entropy flow $Q[0,T]$~\cite{Sekimoto1998,Seifert2005}. Using the moving ISS $\pi_t$ as a baseline, $\Delta\stot$ admits the general trajectory-level decomposition $\Delta\stot = \Delta\srel + \Delta\sact + Y_{\rm HS}$, whose ensemble average recovers the time-integrated form of Eq.~\eqref{eq:macroscopic_split_main}. Here, $\Delta\srel = -\log[\rho_T(X_T)/\pi_T(X_T)] + \log[\rho_0(X_0)/\pi_0(X_0)]$ captures endpoint state relaxation, while $\Delta\sact[0,t]$ accumulates the local active affinity $f_t^{\rm loc}$ along continuous displacements and the non-local affinity $f_t^{\rm nl}$ across individual jumps [defined in Eqs.~\eqref{eq:force_decomp_main} and \eqref{eq:sact_def_main} in the End Matter], satisfying $\dot S_{\rm act}(t)=\frac{\dd}{\dd t}\langle\Delta\sact[0,t]\rangle$ and reducing to standard housekeeping dissipation for autonomous dynamics~\cite{Seifert2012,Esposito2010}. In the Gibbs--ISS setting where $Y_{\rm HS} = W - \Delta F$, this trajectory balance directly reduces to Eq.~\eqref{eq:traj_decomp_intro}.

To isolate the active contribution $\Delta\sact$ at the path-measure level, we construct conjugate dual dynamics with path measure $\PP^\pi$ and generator $\LLop_t^\pi$ that preserve the frozen ISS density $\pi_t$ while reversing its stationary currents~\cite{HatanoSasa2001,Chernyak2006,Esposito2010,ZhangLu2025}:
\begin{equation}
    b_t^\pi=-b_t+2A\nabla\log\pi_t,\qquad
    q_t^\pi(x,y)=\frac{\pi_t(y)q_t(y,x)}{\pi_t(x)}.
  \label{eq:dual_coefficients_main}
\end{equation}
At each fixed protocol value, the forward and dual generators share the same invariant measure,
\begin{equation}
    \LLop_t^*\pi_t=(\LLop_t^\pi)^*\pi_t=0,
  \label{eq:dual_stationarity_main}
\end{equation}
while the dual dynamics reverse both the continuous diffusive current [Eq.~\eqref{eq:dual_continuous_current}] and the directional jump fluxes via $\pi_t(x)q_t^\pi(x,y)=\pi_t(y)q_t(y,x)$ (the dual and forward dynamics coincide under detailed balance). With both processes initialized in $\rho_0$, substituting the dual coefficients~\eqref{eq:dual_coefficients_main} into Girsanov's formula~\eqref{eq:em-relative-entropy} and using $\LLop_t^*\pi_t=0$ gives $\dd\PP^\pi/\dd\PP=\ee^{-\Delta\sact}$ and $\mathcal I^\pi=\Delta\sact$ [End Matter, Eq.~\eqref{eq:em-dual-likelihood}].

Capturing total entropy production and work relations further requires distinguishing direct path reversal from thermodynamic protocol reversal. The path-reversal operator $\Theta$, defined by $(\Theta X)_t=X_{T-t}$ (with standard left-limit conventions at jumps), maps the forward ensemble $\PP$ to the time-reversed measure $\Theta_\#\PP$ initialized in $\rho_T$~\cite{ConfortiLeonard2022,CattiauxConfortiGentilLeonard2023}. Its effective generator retains the diffusion tensor $A$ with drift and jump coefficients governed by the evolving density $\rho_{T-t}$:
\begin{align}
    b_t^\Theta(x)&=-b_{T-t}(x)+2A\nabla\log\rho_{T-t}(x),\nonumber\\
    q_t^\Theta(x,y)&=\frac{\rho_{T-t}(y)q_{T-t}(y,x)}{\rho_{T-t}(x)}.
  \label{eq:operational_theta_jump}
\end{align}
The relative action $\mathcal I^\Theta$ retains the escape-rate correction [Eq.~\eqref{eq:em-relative-entropy}] and depends on the observation time $T$ through the coefficients in Eq.~\eqref{eq:operational_theta_jump}, which use the actual density $\rho_{T-t}$ rather than the frozen ISS $\pi_t$. In contrast, a physical protocol-reversal experiment $R$ drives the system under the time-reversed schedule $b_t^R=b_{T-t}$ and $q_t^R=q_{T-t}$ from an initial distribution $\rho_T$, generating a distinct path measure $\PP_R$~\cite{Seifert2012}. Comparing forward trajectories against the reversed records of this reverse experiment ($\Theta_\#\PP_R$) eliminates the escape-rate mismatch and recovers the entropy production $\Delta\stot$. In the reset example in the End Matter, $\langle\Delta\stot\rangle_\PP=0$ for every $T$, whereas $\langle\mathcal I^\Theta\rangle_\PP$ varies with the observation time~\cite{Fuchs2016,PalRahav2017}. Table~\ref{tab:reference-comparisons} summarizes these three fundamental path comparisons (verified in the End Matter). The same path-measure construction also recovers the Hatano--Sasa equality for an initial steady-state distribution and detailed entropy fluctuation relations between paired forward and conjugate ensembles~\cite{HatanoSasa2001,Esposito2010}.

\begin{table}[htbp]
  \centering
  \caption{Path-measure comparisons on $[0,T]$ and associated trajectory functionals $\mathcal I^\alpha=\log(\dd\PP/\dd\PP^\alpha)$. Forward ($\PP$) and dual ($\PP^\pi$) processes are initialized in $\rho_0$, whereas path-reversed ($\Theta_\#\PP$) and protocol-reversed ($\PP_R$) ensembles are initialized in $\rho_T$. Each comparison satisfies $\langle \ee^{-\mathcal I^\alpha}\rangle_\PP=1$.}
  \label{tab:reference-comparisons}
  \small
  \begin{tabular}{@{}p{0.38\columnwidth}p{0.24\columnwidth}p{0.34\columnwidth}@{}}
    \hline\hline
    Reference dynamics & Measure $\PP^\alpha$ & Relative action $\mathcal I^\alpha$\\
    \hline
    Current-reversed dual & $\PP^\pi$ & $\Delta\sact$\\[3pt]
    Direct path reversal & $\Theta_\#\PP$ & $\mathcal I^\Theta$\\[3pt]
    Protocol-and-path \newline reversal & $\Theta_\#\PP_R$ & $\Delta\stot$\\
    \hline\hline
  \end{tabular}
\end{table}

\paragraph*{Active Jarzynski and Crooks relations.}
For work protocols in the Gibbs--ISS setting $\pi_t(x)=Z_t^{-1}\ee^{-H(t,x)}$, mechanical work is $W=\int_0^T(\partial_t H)(t,X_t)\dd t$ and the equilibrium free-energy difference is $\Delta F=\mathcal F_T-\mathcal F_0=-\log(Z_T/Z_0)$. We initialize the forward ($\PP$) and dual ($\PP^\pi$) processes in $\pi_0$, and the protocol-reversed experiment ($\PP_R$) in $\pi_T$. By the trajectory decomposition in Eq.~\eqref{eq:traj_decomp_intro}, $W+\Delta\sact-\Delta F=\Delta\stot-\Delta\srel$. Since the dual relative action $\mathcal I^\pi=\Delta\sact$ captures the active term in this balance, applying the exponential work tilt $\Gamma=W$ to $\PP^\pi$ via Eq.~\eqref{eq:gauge_path_measure} defines $\dd\PP^{\pi,W}=(Z_\pi^W)^{-1}\ee^{-W}\dd\PP^\pi$, so that changing measure via Eq.~\eqref{eq:gauge_chain_main} gives
\begin{equation}
    \langle e^{-(\Delta\sact+W)}\rangle_{\PP}
    = \langle e^{-W}\rangle_{\PP^\pi}=Z_\pi^W.
  \label{eq:active_expectation_main}
\end{equation}
To evaluate $Z_\pi^W$, we use the stationarity of the Gibbs density under the frozen dual dynamics and the time derivative of the Gibbs factor:
\begin{equation}
    (\LLop_t^\pi)^*\ee^{-H(t,\cdot)}=0,\quad
    \partial_t \ee^{-H(t,x)}=-(\partial_tH)(t,x)\ee^{-H(t,x)}.
  \label{eq:dual_gibbs_work_main}
\end{equation}
Applying Eq.~\eqref{eq:dual_gibbs_work_main} to the Feynman--Kac equation~\eqref{eq:em-gauge-pde} with $K_t=\LLop_t^\pi$ gives $Z_\pi^W=Z_T/Z_0=\ee^{-\Delta F}$ through the weighted endpoint identity~\eqref{eq:em-weighted-work} in the End Matter. Combining this value of the normalization factor with the reverse-path comparison in Eq.~\eqref{eq:em-reverse-work-likelihood} gives the measure identity
\begin{equation}
    \frac{\dd\PP^{\pi,W}}{\dd\PP}
    =\ee^{-(W+\Delta\sact-\Delta F)}
    =\frac{\dd(\Theta_\#\PP_R)}{\dd\PP}.
  \label{eq:active_weight_main}
\end{equation}
Thus $\PP^{\pi,W}=\Theta_\#\PP_R$: work-tilting the dual process reproduces the path-reversed ensemble of the protocol-reversal experiment. Integrating Eq.~\eqref{eq:active_weight_main} over all forward trajectories yields the Active Jarzynski equality:
\begin{equation}
    \left\langle
    \ee^{-(W+\Delta\sact)}
    \right\rangle_{\pi_0}=\ee^{-\Delta F}.
  \label{eq:active_jarzynski}
\end{equation}
Because the forward generator similarly satisfies $\LLop_t^*\ee^{-H(t,\cdot)}=0$, applying $\Gamma=W$ directly to the forward reference $\PP^\alpha=\PP$ ($\mathcal{I}^\alpha=0$) yields the conventional mechanical Jarzynski equality $\langle \ee^{-W}\rangle_{\pi_0}=\ee^{-\Delta F}$~\cite{Jarzynski1997PRL,Hartmann2019}, which coincides with Eq.~\eqref{eq:active_jarzynski} when $\Delta\sact$ vanishes.

When the work distributions admit densities, setting $\mathcal O=W$ in Eq.~\eqref{eq:conditional_observable} for a reference measure $\PP^\alpha$ gives the density ratio
\begin{equation}
    \frac{p_F(w)}{p^\alpha(w)}
    =\left[\langle e^{-\mathcal I^\alpha}\mid W=w\rangle_{\PP}\right]^{-1},
  \label{eq:conditional_work_main}
\end{equation}
where $p_F$ and $p^\alpha$ are the work densities under $\PP$ and $\PP^\alpha$, respectively. Under the reversed protocol $H^R(t,x)=H(T-t,x)$, work is odd under path reversal, $W^R(\Theta X)=-W(X)$, so choosing $\PP^\alpha=\Theta_\#\PP_R$ yields $p^\alpha(w)=p_R(-w)$. Substituting the likelihood ratio from Eq.~\eqref{eq:active_weight_main} into Eq.~\eqref{eq:conditional_work_main} then establishes the Active Crooks relation:
\begin{equation}
    \frac{p_F(w)}{p_R(-w)}
    =\frac{e^{w-\Delta F}}
    {\langle e^{-\Delta\sact}\mid W=w\rangle_{\PP}}.
  \label{eq:crooks_general_main}
\end{equation}
Physically, forward trajectories with identical mechanical work $w$ can accumulate different active entropies $\Delta\sact$ and thus carry distinct likelihood ratios relative to their time-reversed counterparts. The conditional factor in Eq.~\eqref{eq:crooks_general_main} averages these active weights over the constant-work subensemble. It also equals the ratio of the dual to the forward work density at the same $w$ [Eq.~\eqref{eq:em-dual-work-density} in the End Matter], quantifying how reversing stationary active currents alters work statistics. Solving Eq.~\eqref{eq:crooks_general_main} for $p_R(-w)$ and integrating over $w$ yields the Active Jarzynski equality [Eq.~\eqref{eq:active_jarzynski}] by the law of total expectation, while vanishing active affinities ($f_t^{\rm loc}=f_t^{\rm nl}=0$) reduce Eq.~\eqref{eq:crooks_general_main} to the classical Crooks relation $p_F(w)/p_R(-w)=\ee^{w-\Delta F}$~\cite{Crooks1999}. Table~\ref{tab:ft-applications} summarizes these three work relations.

\begin{table}[htbp]
  \centering
  \caption{Work relations for the Gibbsian ISS $\pi_t(x)=Z_t^{-1}\ee^{-H(t,x)}$ with $\Gamma=W$. Forward ($\PP$) and dual ($\PP^\pi$) processes start from $\pi_0$, and the protocol-reversed process ($\PP_R$) starts from $\pi_T$. For both active relations, the work-tilted dual satisfies $\PP^{\pi,W}=\Theta_\#\PP_R$.}
  \label{tab:ft-applications}
  \small
  \begin{tabular}{@{}p{0.36\columnwidth}p{0.28\columnwidth}p{0.32\columnwidth}@{}}
    \hline\hline
    Work relation & {\raggedright Reference before weighting\par} & Trajectory domain\\
    \hline
    Mechanical Jarzynski & $\PP$ ($\mathcal I^\alpha=0$) & Full path space\\[3pt]
    Active Jarzynski & $\PP^\pi$ ($\mathcal I^\alpha=\Delta\sact$) & Full path space\\[3pt]
    Active Crooks & $\PP^\pi$ ($\mathcal I^\alpha=\Delta\sact$) & Fixed-work subensemble\\
    \hline\hline
  \end{tabular}
\end{table}

\paragraph*{Examples.}

We illustrate the coexistence of active dissipation and the recovery of free-energy differences with a two-dimensional overdamped particle subject to chiral active drift and Poisson jumps in the harmonic potential $H(t,x)=\lambda_t\vert{}x\vert{}^2/2$ with time-dependent stiffness $\lambda_t$. Its position $X_t \in \mathbb{R}^2$ obeys
\begin{align}
    \dd X_t={}&\lambda_t(\mathsf R_{\mathrm{rot}}-I_2)X_t\dd t
    +\sqrt2\,\dd W_t\nonumber\\
    &+\int_{\mathbb R^2}(y-X_{t^-})N_X(\dd t,\dd y),
  \label{eq:shared-jump-diffusion}
\end{align}
where $\mathsf R_{\mathrm{rot}}x=(-x_2,x_1)$ is a $90^\circ$ rotation. Jumps occur at rate one. Conditional on $X_{t^-}=x$, the new position has distribution $\mathcal N(\mathsf R_{\mathrm{rot}}x/2,3I_2/(4\lambda_t))$. Both the diffusion and the jumps preserve the frozen density $\pi_t(x)=\lambda_t e^{-\lambda_t|x|^2/2}/(2\pi)=e^{-H(t,x)}/Z_t$. We initialize with $X_0\sim\mathcal N(0,I_2)$. Here, $t$ denotes the evolution time. We evaluate all observables over $[0,T]$, with $T\in[0,1]$. The free energy $\mathcal F_T$ is defined by the partition function of $H(T,\cdot)$; the evolving density need not equal $\pi_T$. The End Matter gives the numerical method for both protocols.

\begin{figure}[htbp]
  \centering
  \centering
  \includegraphics[width=\columnwidth]{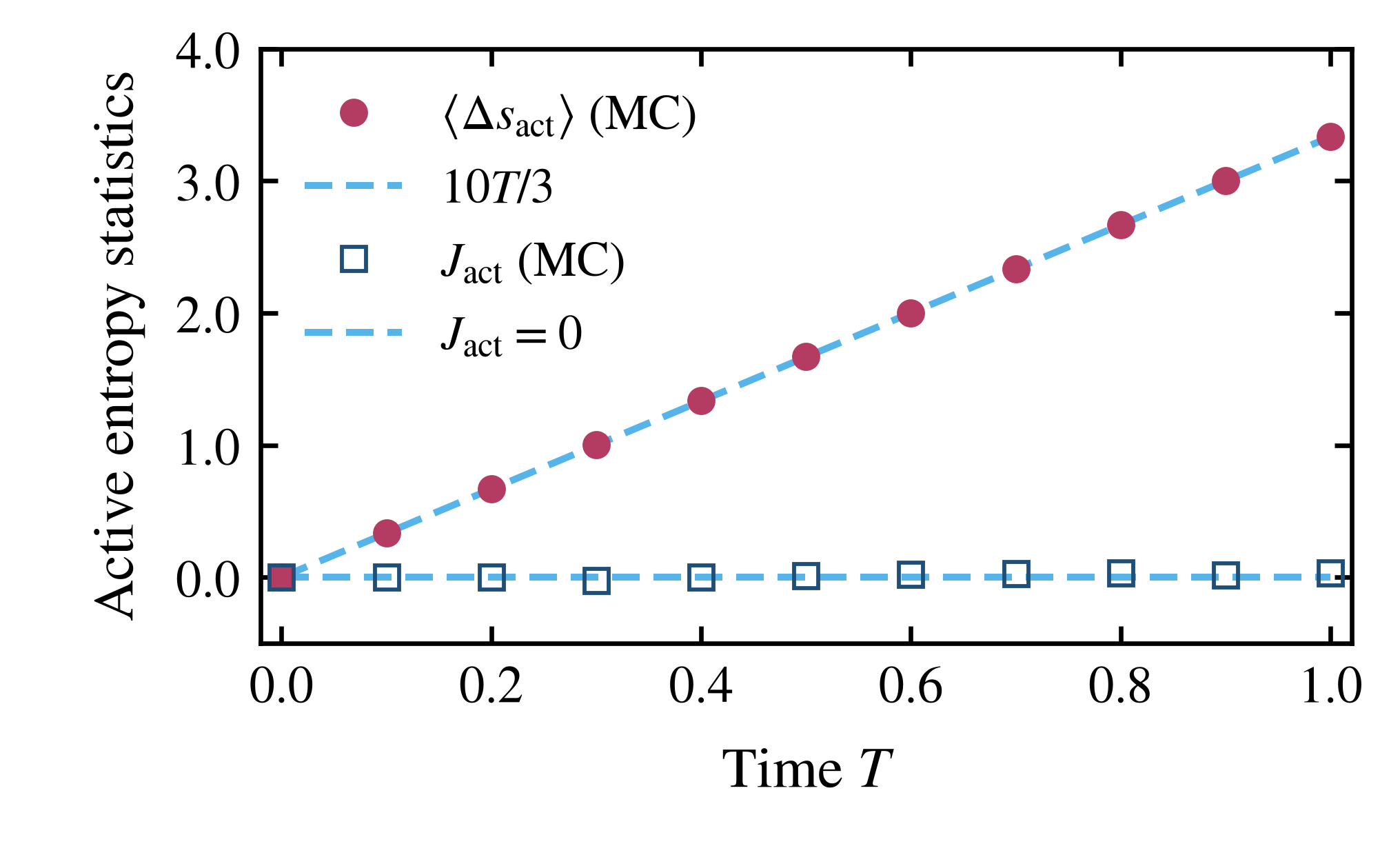}
  \caption{Active statistics at fixed landscape. The sample mean of $\Delta s_{\rm act}$ is compared with $10T/3$. Here $J_{\rm act}(T)=-\log\langle e^{-\Delta s_{\rm act}[0,T]}\rangle_{\pi_0}=0$; the open squares show its Monte Carlo estimates.}
  \label{fig:active_storage}

\end{figure}

For the fixed landscape $\lambda_t=1$, the actual density remains $\rho_t=\pi$. Rotational drift and asymmetric jumps produce active entropy without protocol work or free-energy change:
\begin{gather}
    W=\Delta F=0,\nonumber\\
    \langle\Delta s_{\rm act}[0,T]\rangle=\frac{10T}{3},
    \qquad \langle e^{-\Delta s_{\rm act}[0,T]}\rangle=1.
  \label{eq:fixed-active-main}
\end{gather}
Thus active entropy accumulates without a change in either the state distribution or free energy (Fig.~\ref{fig:active_storage}). The positive mean and the unit exponential average describe different statistics of the same active trajectories.

\begin{figure}[htbp]
  \centering
  \centering
  \includegraphics[width=\columnwidth]{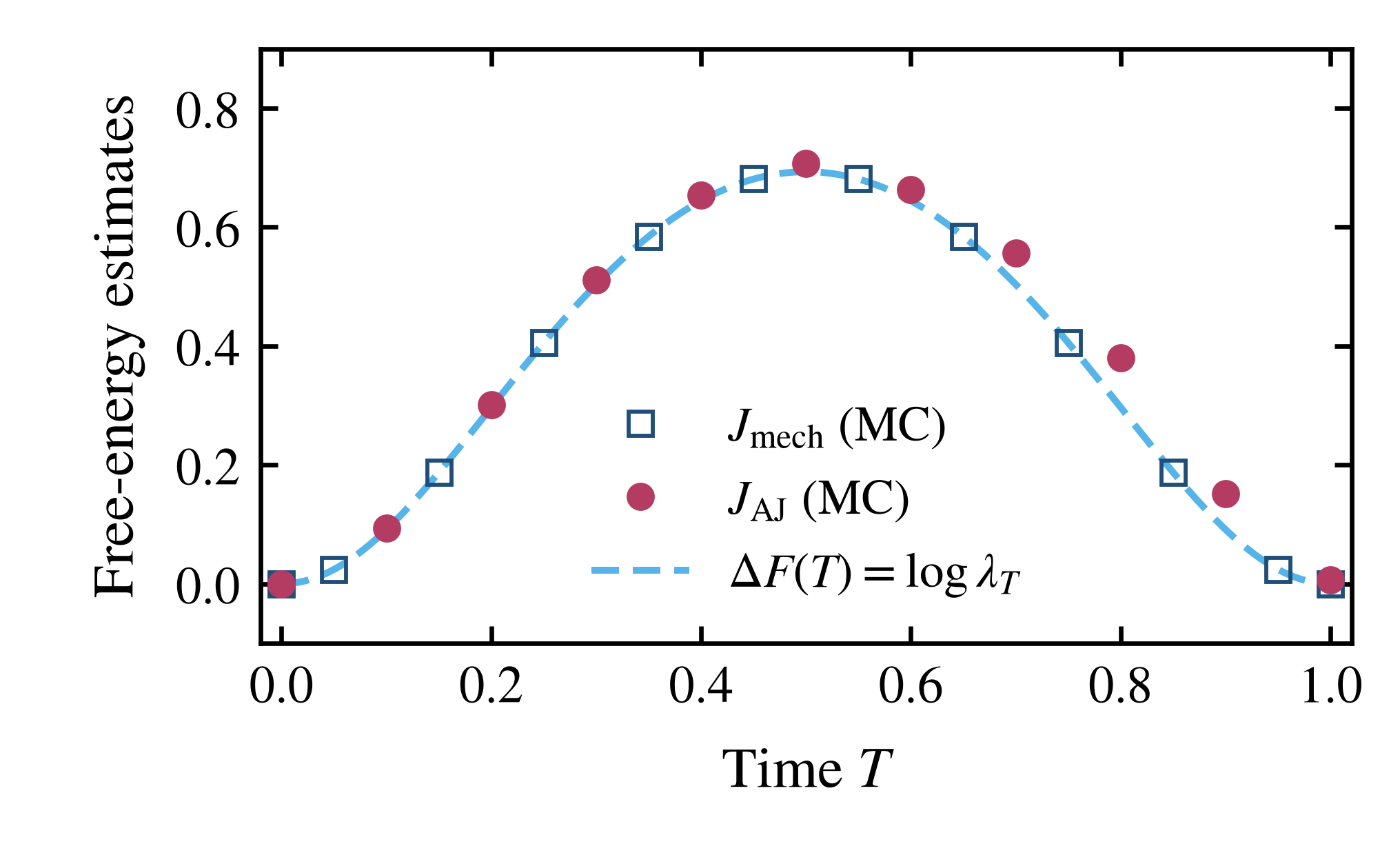}
  \caption{Mechanical and Active Jarzynski estimates for $\lambda_t=\tfrac32-\tfrac12\cos(2\pi t)$. Open squares and filled circles show Monte Carlo estimates of $J_{\rm mech}(T)$ and $J_{\rm AJ}(T)$, respectively. The dashed curve gives their common theoretical value $\Delta F_T=\log\lambda_T$.}
  \label{fig:active_jarzynski}

\end{figure}

Keeping the same model, we apply the periodic stiffness protocol $\lambda_t=\tfrac32-\tfrac12\cos(2\pi t)$. The trap compresses during the first half-cycle and expands during the second. The mechanical work and Gibbs free-energy difference are
\begin{equation}
    W[0,T]=\frac12\int_0^T\dot\lambda_t|X_t|^2\dd t,
    \qquad \Delta F_T=\log\lambda_T.
\end{equation}
The Gibbsian ISS $\pi_t$ ensures that both the mechanical and Active Jarzynski relations hold for this protocol:
\begin{align}
    J_{\rm mech}(T)&:=-\log\langle e^{-W}\rangle_{\pi_0}
    =\Delta F_T,\\
    J_{\rm AJ}(T)&:=-\log\langle e^{-(W+\Delta\sact)}\rangle_{\pi_0}
    =\Delta F_T.
\end{align}
Fig.~\ref{fig:active_jarzynski} compares their sample estimates. At $T=1$ the Hamiltonian returns to its initial value and $\Delta F_1=0$. Since work fluctuates across trajectories, strict Jensen's inequality applied to $\langle e^{-W}\rangle_{\pi_0}=1$ gives $\langle W\rangle_{\pi_0}>0$, while nonzero active currents yield positive mean active entropy. A spatial reflection $M=\operatorname{diag}(1,-1)$ maps the original dynamics to their dual while preserving $|X_t|^2$, and hence $W$. Consequently, $\langle e^{-\Delta\sact}\mid W=w\rangle_{\PP}=1$, so this model has the classical Crooks work ratio despite positive mean active entropy.

\paragraph*{Discussion.}
By combining current-reversed dual dynamics with exponential gauge tilts, we have established a general framework for finite-time fluctuation relations in driven jump-diffusions without detailed balance. A central physical insight of our work is the coexistence of the mechanical and Active Jarzynski equalities in systems admitting a Gibbsian ISS: while the conventional exponential work average $\langle e^{-W}\rangle$ recovers the free-energy difference using the stationarity of the Gibbs density under the frozen forward dynamics, the active counterpart $\langle e^{-(W+\Delta s_{\rm act})}\rangle$ incorporates the housekeeping heat associated with the non-equilibrium currents using the corresponding stationarity under the frozen dual dynamics. In the corresponding Active Crooks relation [Eq.~\eqref{eq:crooks_general_main}], the conditional factor $\langle e^{-\Delta s_{\rm act}}\mid W=w\rangle_{\PP}$ measures how autonomous active fluctuations skew the work distribution away from the classical Crooks work relation. Beyond the Gibbs--ISS regime, our general weighted endpoint identity (detailed in the End Matter) remains valid for arbitrary reference Hamiltonians. Experimentally, our results are directly applicable to optically trapped active colloids, engineered microswimmers with run-and-tumble or Lévy-flight kinematics, and single-molecule manipulation in active viscoelastic media, offering a rigorous route to simultaneously reconstruct effective free-energy landscapes and quantify hidden housekeeping dissipation from finite-time trajectory measurements.
\paragraph*{Acknowledgments.}
We thank Liu Hong, Qiao Huang, Yubin Lu, and Henri Orland for helpful discussions. S.F. and J.D. acknowledge support from the Guangdong Provincial Key Laboratory of Mathematical and Neural Dynamical Systems (Grant No. 2024B1212010004), the Dongguan Key Laboratory for Data Science and Intelligent Medicine, the Innovation Team Project for Higher Education Institutions in Guangdong Province, ``Cross-disciplinary Research Team on Data Science and Intelligent Medicine'' (Grant No. 2023KCXTD054), the International Collaboration Fund for Creative Research Teams of the National Natural Science Foundation of China (Grant No. W2541005), and the Natural Science Foundation of Guangdong Province (Grant No. 2025A1515011188). B.M. acknowledges support from the National Natural Science Foundation of China (NSFC) (Grant No. 12575045). We used GPT to assist with checking and improving the draft, and DeepSeek and GLM to assist with the numerical simulations.

\clearpage
\raggedbottom
\begingroup

\section*{End Matter}
The compared path measures, including their initial distributions, satisfy $\PP^\alpha\sim\PP$, and their Girsanov density processes are assumed to be true martingales of unit expectation \cite{CheriditoFilipovicYor2005,JacodShiryaev2003}. We retain the main-text notation, with all state variables even under reversal and $X^c$ denoting the continuous component, $\dd X_t^c=b_t(X_t)\dd t+\sqrt2a\,\dd W_t$. We write $t_j$ for the jump times and use $\partial_t$ for time differentiation at fixed position.

\paragraph*{Currents and entropy production.}
The Fokker--Planck equation is
\begin{equation*}
    \partial_t\rho_t=-\nabla\cdot\Jloc_t
    -\int\Jnl_t(x,y)\dd y,
\end{equation*}
with currents
\begin{align*}
    \Jloc_t&=b_t\rho_t-A\nabla\rho_t,\\
    \Jnl_t(x,y)&=\rho_t(x)q_t(x,y)-\rho_t(y)q_t(y,x).
\end{align*}
For the ISS $\pi_t$, we define the active affinities
\begin{align}
    f_t^{\rm loc}(x)&=A^{-1}b_t(x)-\nabla \log\pi_t(x),\nonumber\\
    f_t^{\rm nl}(x,y)&=\log
    \frac{\pi_t(x)q_t(x,y)}{\pi_t(y)q_t(y,x)}.
  \label{eq:force_decomp_main}
\end{align}
The total thermodynamic affinities satisfy
\begin{align*}
    A^{-1}\frac{\Jloc_t}{\rho_t}
    &=f_t^{\rm loc}-\nabla\log\frac{\rho_t}{\pi_t},\\
    \log\frac{\rho_t(x)q_t(x,y)}{\rho_t(y)q_t(y,x)}
    &=f_t^{\rm nl}(x,y)+\log\frac{\rho_t(x)}{\pi_t(x)}
    -\log\frac{\rho_t(y)}{\pi_t(y)}.
\end{align*}
At each frozen time $t$, $f_t^{\rm loc}$ is the continuous active force and $f_t^{\rm nl}$ the jump affinity sustaining stationary currents. The terms involving $\rho_t/\pi_t$ drive relaxation toward $\pi_t$ under the frozen dynamics and give the nonadiabatic entropy-production rate when paired with the corresponding currents~\cite{Esposito2010}.
The total entropy-production rate has the current--affinity form
\begin{align}
    \dot S_{\mathrm{tot}}(t)
    ={}&\int\Jloc_t\cdot A^{-1}\frac{\Jloc_t}{\rho_t}\dd x\nonumber\\
    &+\frac12\iint\Jnl_t(x,y)
    \log\frac{\rho_t(x)q_t(x,y)}{\rho_t(y)q_t(y,x)}\dd x\dd y.
  \label{eq:em-total-rate}
\end{align}
Substituting the two affinity decompositions above into Eq.~\eqref{eq:em-total-rate} gives the active and nonadiabatic rates $\dot S_{\rm act}(t)$ and $\dot S_{\rm na}(t)$ in Eq.~\eqref{eq:macroscopic_split_main},
\begin{align}
    \dot S_{\mathrm{act}}(t)
    ={}&\int\Jloc_t\cdot f_t^{\rm loc}\dd x
    +\frac12\iint\Jnl_t f_t^{\rm nl}\dd x\dd y,
  \label{eq:em-active-rate}\\
    \dot S_{\mathrm{na}}(t)
    ={}&\dot S_{\rm tot}(t)-\dot S_{\rm act}(t)=-\int(\LLop_t^*\rho_t)\log\frac{\rho_t}{\pi_t}\dd x.
  \label{eq:em-na-rate}
\end{align}
The last equality uses integration by parts and antisymmetry of $\Jnl_t$. Differentiating $D(\rho_t\Vert\pi_t)$ with respect to $t$ yields Eq.~\eqref{eq:nonadiabatic_rate_main}. The total rate is nonnegative because its continuous term is quadratic and $(u-v)\log(u/v)\ge0$. For $h=\rho_t/\pi_t$, concavity of the logarithm gives $h\LLop_t\log h\le\LLop_t h$, while $\int\pi_t\LLop_t h\dd x=0$. Thus $\dot S_{\rm na}=-\int\pi_t h\LLop_t\log h\dd x\ge0$.

\paragraph*{Path likelihoods.}
For reference coefficients $b_t^\alpha,q_t^\alpha$ and initial density $\rho_0^\alpha$, write $u_t=(\sqrt2\,a)^{-1}(b_t^\alpha-b_t)$. Girsanov's formula along the forward trajectory gives
\begin{align}
    \mathcal I^\alpha
    ={}&\log\frac{\rho_0(X_0)}{\rho_0^\alpha(X_0)}\nonumber\\
    &-\int_0^T u_t(X_t)\cdot\dd W_t
    +\frac12\int_0^T|u_t(X_t)|^2\dd t\nonumber\\
    &+\sum_{t_j\le T}\log
    \frac{q_{t_j}(X_{t_j^-},X_{t_j})}
    {q_{t_j}^\alpha(X_{t_j^-},X_{t_j})}\nonumber\\
    &-\int_0^T\!\!\int[q_t-q_t^\alpha](X_t,y)\dd y\dd t .
  \label{eq:em-relative-entropy}
\end{align}
In the $\PP$-expectation of Eq.~\eqref{eq:em-relative-entropy}, the Brownian integral vanishes and the jump sum is evaluated using the compensator $q_t(X_{t^-},y)\dd y\dd t$, giving Eq.~\eqref{eq:om_mean_main}. The escape-rate correction integrates $q_t-q_t^\alpha$ as a single difference. For $\Theta_\#\PP$, the initial contribution is $\log[\rho_0(X_0)/\rho_T(X_0)]$, with mean $D(\rho_0\Vert\rho_T)$.

\paragraph*{ISS decomposition of entropy production.}
At the frozen density $\pi_t$, the dual coefficients give
\begin{equation}
    b_t^\pi\pi_t-A\nabla\pi_t
    =-\bigl[b_t\pi_t-A\nabla\pi_t\bigr].
  \label{eq:dual_continuous_current}
\end{equation}
The environment-entropy flow in the entropy balance is
\begin{equation*}
    Q[0,T]=\int_0^T A^{-1}b_t(X_t)\circ\dd X_t^c
    +\sum_{t_j\le T}\log
    \frac{q_{t_j}(X_{t_j^-},X_{t_j})}
    {q_{t_j}(X_{t_j},X_{t_j^-})}.
\end{equation*}
Here $\circ$ denotes Stratonovich integration. Using the affinities in Eq.~\eqref{eq:force_decomp_main}, the active contribution is
\begin{equation}
    \Delta\sact[0,T]
    =\int_0^Tf_t^{\rm loc}(X_t)\circ\dd X_t^c
    +\sum_{t_j\le T}f_{t_j}^{\rm nl}(X_{t_j^-},X_{t_j}).
  \label{eq:sact_def_main}
\end{equation}
The continuous term integrates the active force along displacements, and each jump contributes $f_t^{\rm nl}(x,y)$. Set $\Vss_t=-\log\pi_t$. Substituting Eq.~\eqref{eq:force_decomp_main} into Eq.~\eqref{eq:sact_def_main} and applying the chain rule to $\Vss_t(X_t)$ gives $\Delta\sact=Q+\Vss_T(X_T)-\Vss_0(X_0)-Y_{\rm HS}$. Using the main-text definitions of $\Delta s_{\rm sys}$ and $\Delta\srel$ recovers the trajectory entropy balance. To evaluate the $\PP$-expectations of $\Delta\stot=\Delta s_{\rm sys}+Q$ and Eq.~\eqref{eq:sact_def_main}, we use Stratonovich-to-It\^o conversion and the jump compensator. Differentiating these expectations with respect to $T$ gives $\dot S_{\rm tot}(T)$ and $\dot S_{\rm act}(T)$ in Eqs.~\eqref{eq:em-total-rate} and \eqref{eq:em-active-rate}. Averaging the trajectory entropy balance and differentiating with respect to $T$ gives Eq.~\eqref{eq:macroscopic_split_main}. For a Gibbsian ISS, substituting $\Vss_t(x)=H(t,x)-\mathcal F_t$ and $Y_{\rm HS}=W-\Delta F$ into the trajectory identity for $\Delta\sact$ gives
\begin{equation}
    Q+H(T,X_T)-H(0,X_0)=W+\Delta\sact.
  \label{eq:energy_work_main}
\end{equation}
For transitions mediated by a single heat bath, the Einstein relation for diffusion and local detailed balance for jumps identify $Q$ as the heat released to the bath in thermal units \cite{Sekimoto1998,Maes2021}. Equation~\eqref{eq:energy_work_main} then expresses the first law, with supplied work $W+\Delta\sact$.

For forward and dual processes with the same initial density, substitution of Eq.~\eqref{eq:dual_coefficients_main} into Eq.~\eqref{eq:em-relative-entropy} combines the continuous drift correction and the jump escape-rate correction into $\LLop_t^*\pi_t/\pi_t$. Hence
\begin{equation}
    \mathcal I^\pi=\Delta\sact+\int_0^T\frac{\LLop_t^*\pi_t}{\pi_t}(X_t)\dd t=\Delta\sact.
  \label{eq:em-dual-likelihood}
\end{equation}
The rate of path-space relative entropy in Eq.~\eqref{eq:om_mean_main} is nonnegative for this comparison, giving $\dot S_{\rm act}(t)\ge0$.

For a protocol-reversed experiment with initial density $\rho_0^R$, let $\rho_s^R$ denote its density at time $s$. The reversed-time density satisfies $\partial_t(\rho_{T-t}^R)+\LLop_t^*\rho_{T-t}^R=0$. Applying Eq.~\eqref{eq:em-relative-entropy} with the path-reversal coefficients built from $\rho_{T-t}^R$, followed by the chain rule for $\log\rho_{T-t}^R(X_t)$, gives
\begin{align}
    \log\frac{\dd\PP}{\dd(\Theta_\#\PP_R)}
    ={}&\log\frac{\rho_0(X_0)}{\rho_0^R(X_T)}+Q[0,T]\nonumber\\
    &+\int_0^T\frac{\partial_t(\rho_{T-t}^R)+\LLop_t^*\rho_{T-t}^R}{\rho_{T-t}^R}(X_t)\dd t.
  \label{eq:em-reverse-likelihood}
\end{align}
The integral vanishes by the equation for $\rho_{T-t}^R$. Taking $\rho_0^R=\rho_T$ gives $\Delta\stot$, as in Table~\ref{tab:reference-comparisons}. With forward and protocol-reversed initial densities $\pi_0$ and $\pi_T$, respectively, the same formula gives
\begin{align}
    \log\frac{\dd\PP}{\dd(\Theta_\#\PP_R)}
    &=\log\frac{\pi_0(X_0)}{\pi_T(X_T)}+Q[0,T]
  \label{eq:em-reverse-work-likelihood}\\
    &=H(T,X_T)-H(0,X_0)+Q[0,T]-\Delta F.
    \nonumber
\end{align}
The second line uses the Gibbs densities. Substituting $Q+H(T,X_T)-H(0,X_0)=W+\Delta\sact$ from Eq.~\eqref{eq:energy_work_main} into Eq.~\eqref{eq:em-reverse-work-likelihood}, then exponentiating the negative of this log ratio, gives the path-measure likelihood ratio in Eq.~\eqref{eq:active_weight_main}.

\paragraph*{Weighted endpoint identity.} For a prescribed Hamiltonian $H(t,x)$, initialize the original process from $e^{-H(0,\cdot)}/Z_0$. For a smooth bounded test function $\varphi$, we evaluate $\langle\varphi(X_T)e^{-\Gamma[0,T]}\rangle_\PP$, where
\begin{equation*}
    \Gamma[0,T]=Q[0,T]+H(T,X_T)-H(0,X_0).
\end{equation*}
When the Gibbs density $\pi_t=e^{-H(t,\cdot)}/Z_t$ satisfies $\LLop_t^*\pi_t=0$, Eq.~\eqref{eq:energy_work_main} gives $\Gamma=W+\Delta\sact$. For the dual initialized from the same density $\pi_0$, Eq.~\eqref{eq:em-dual-likelihood} then gives $e^{-\Gamma}\dd\PP=e^{-W}\dd\PP^\pi$. For general $H(t,x)$, set $g(t,x)=\langle\varphi(X_T)\ee^{-\Gamma[t,T]}\rangle_{t,x}$, where $\langle\cdot\rangle_{t,x}$ refers to the original process started at $(t,x)$. By the definition of $g$, $e^{-\Gamma[t,s]}g(s,X_s)$ is a martingale in $s$, so its It\^o drift vanishes, giving
\begin{equation}
    \partial_tg+K_tg-(\partial_tH)(t,x)g=0,\qquad g(T,x)=\varphi(x).
  \label{eq:em-gauge-pde}
\end{equation}
The tilted spatial operator $K_t$, generally with $K_t1\ne0$, and its Lebesgue adjoint satisfy
\begin{equation*}
  \begin{aligned}
    K_t\varphi&=\ee^{H(t,x)}\LLop_t^*(\ee^{-H(t,x)}\varphi),\\
    K_t^*\ee^{-H(t,x)}&=\ee^{-H(t,x)}\LLop_t1=0.
  \end{aligned}
\end{equation*}
For $I(t)=\int e^{-H(t,x)}g(t,x)\dd x$, differentiation with respect to $t$ using Eq.~\eqref{eq:em-gauge-pde} cancels the terms containing $\partial_tH$. Integration by parts then gives $I'(t)=-\int gK_t^*e^{-H(t,\cdot)}\dd x=0$. Evaluating $I(0)=I(T)$ and dividing by $Z_0$ yields
\begin{equation}
    \bigl\langle\varphi(X_T)\ee^{-\Gamma[0,T]}\bigr\rangle_{\ee^{-H(0,\cdot)}/Z_0}
    =\frac1{Z_0}\int\varphi(x)\ee^{-H(T,x)}\dd x.
  \label{eq:em-weighted-work}
\end{equation}
Equation~\eqref{eq:em-reverse-likelihood} with reverse initial density $e^{-H(T,\cdot)}/Z_T$ gives $e^{-\Gamma}\dd\PP=(Z_T/Z_0)\dd(\Theta_\#\PP_R)$, so Eq.~\eqref{eq:em-weighted-work} also holds for bounded measurable $\varphi$. Taking $\varphi=1$ gives $\langle e^{-\Gamma}\rangle_\PP=Z_T/Z_0=e^{-\Delta F}$. When $\pi_t=e^{-H(t,\cdot)}/Z_t$ is the ISS, $K_t=\LLop_t^\pi$ and $e^{-\Gamma}\dd\PP=e^{-W}\dd\PP^\pi$ give $Z_\pi^W=Z_T/Z_0$ and Eq.~\eqref{eq:active_jarzynski}.

For forward and dual processes initialized from $\pi_0$ under the same protocol, conditioning $\dd\PP^\pi/\dd\PP=e^{-\Delta\sact}$ from Eq.~\eqref{eq:em-dual-likelihood} on $W=w$ gives
\begin{equation}
    \left\langle e^{-\Delta\sact}\mid W=w\right\rangle_{\PP}
    =\frac{p^\pi(w)}{p_F(w)}.
  \label{eq:em-dual-work-density}
\end{equation}
Here $p_F$ and $p^\pi$ are the densities of the same work functional $W$ under $\PP$ and $\PP^\pi$, respectively. The equality holds almost everywhere on the support of $p_F$.

\paragraph*{Examples and numerical methods.}
For each protocol of Eq.~\eqref{eq:shared-jump-diffusion}, we use jump-adapted Euler--Maruyama \cite[Chap.~8]{PlatenBrutiLiberati2010} with a maximum step $10^{-3}$ and a total of $5\times10^6$ trajectories from five independent seeds. Jump waiting times are sampled from an exponential distribution with unit rate. Observables are recorded at intervals of $0.01$ over $[0,1]$. The continuous contribution to the active entropy is evaluated using left-point It\^o sums, since the rotational field has zero Stratonovich correction, and jump affinities are summed separately. Work is computed using trapezoidal quadrature on each jump-free segment, taking the terminal position before any jump. Exponential weights are pooled across all trajectories before taking the negative logarithm.

To compare $\mathcal I^\Theta$ with $\Delta\stot$, take $A=I$, $b=-\nabla V$, $q_t(x,y)=\gamma_t\rho_{\rm eq}(y)$, and $X_0\sim\rho_{\rm eq}$ in Eq.~\eqref{eq:sde_main}, where $\rho_{\rm eq}=Z^{-1}e^{-V}$ \cite{Fuchs2016,PalRahav2017}. These jumps redraw from $\rho_{\rm eq}$ and satisfy detailed balance at each time, unlike unidirectional resetting to a fixed point. Then $\rho_t=\rho_{\rm eq}$ and $q_t^\Theta(x,y)=\gamma_{T-t}\rho_{\rm eq}(y)$. Equations~\eqref{eq:om_mean_main} and \eqref{eq:em-total-rate} give $\langle\mathcal I^\Theta\rangle_\PP=\int_0^T[\gamma_t\log(\gamma_t/\gamma_{T-t})-\gamma_t+\gamma_{T-t}]\dd t$ and $\langle\Delta\stot\rangle_\PP=0$. For $\gamma_t=2+\sin t$, $\langle\mathcal I^\Theta\rangle_\PP=0$ when $T=\pi$, whereas it is positive when $T=2\pi$.
\endgroup

\end{document}